\documentclass[9pt,twocolumn,twoside]{opticajnl}
\journal{opticajournal} % use for journal or Optica Open submissions
\usepackage{graphicx}
\usepackage{subcaption}
\setboolean{shortarticle}{true}
\usepackage{lineno}

\title{Polarization-based Metalenses with High Numerical Aperture and Focusing Efficiency Utilizing Silicon-rich Nitride}

\author[1,$\dag$]{Alireza Khalilian}
\author[1,$\dag$]{Bowen Yu}
\author[1,2,*]{Yasha Yi}

\affil[1]{Integrated Nano Optoelectronics Laboratory, University of Michigan, 4901 Evergreen Rd., Dearborn, Michigan 48128, USA}
\affil[2]{Energy Institute, University of Michigan, 2301 Bonisteel Blvd., Ann Arbor, Michigan 48109-2100, USA}
\affil[$\dag$]{The authors contributed equally to this work.}
\affil[*]{yashayi@umich.edu}

\begin{abstract}
We explore the cutting-edge application of silicon-rich nitride (SRN) in the realm of high numerical aperture (NA) metalens design, focusing on the crucial role of pitch size optimization in amplifying lens efficiency through advanced simulations. Our investigation unveils how the exceptional tunable high refractive index of SRN can be harnessed to achieve significant advancements in metalens performance. By meticulously designing and simulating two innovative SRN-based metalenses — Mk1, with an NA of $0.9$, reaching an impressive $75\%$ focusing efficiency with full width at half maximum (FWHM) of $0.53\lambda$, and Mk2, with an NA of $0.99$, achieving a $42\%$ efficiency while maintaining an FWHM of $0.48\lambda$. We demonstrate the critical influence of reduced pitch size on enhancing efficiency. This study not only highlights the unparalleled potential of SRN in optimizing metalens efficiency but also represents a significant leap forward in the field of nanophotonics, offering new pathways for the development of highly efficient flat photonic devices.
\end{abstract}

\setboolean{displaycopyright}{false} % Do not include copyright or licensing information in submission.

\begin{document}

\maketitle

\section{Introduction}
The advent of metasurfaces, characterized by engineered nano-scale designs, represents a significant shift in lens technology. These compact metasurfaces offer advantages over conventional bulk lenses in controlling key optical parameters like phase, transmission, and polarization. This has opened up a broad spectrum of applications, ranging from achromatic focusing \cite{chen2021modified}, color routing \cite{chen2021nanophotonic}, to polarization separation \cite{ishizuka2023linear}.
Metasurfaces have transformed traditional optics by enabling unprecedented control over light at the nano-scale. The fundamental work by Yu et al. \cite{yu2011light}, which introduced the generalized laws of reflection and refraction for metasurfaces, marked a turning point in the field. Following this, researchers explored the practical applications of metasurfaces. In their pioneering work, Palermo et al. introduced a reconfigurable metalens that combines nanostructured silica with nematic liquid crystal and gold nanoparticles, enabling dynamic focal length adjustment through thermoplasmonic effects \cite{palermo2022all}. This innovation marks a significant advancement in adaptable optical devices. In their study, Wei Ting Chen et al. present a breakthrough in metalens technology by creating an ultra-thin, single-layer achromatic metalens capable of focusing and imaging light across the visible spectrum (470 to 670 nm) \cite{chen2018broadband}. This innovation, achieved through a novel nanofin design on the metalens surface, addresses the challenge of maintaining functionality in nanoscale optical devices without complex layering or multiplexing. Another study by Khorasaninejad et al. \cite{khorasaninejad2016metalenses} introduced metalens design capable of focusing visible light using TiO$_2$. In another study, Liang \cite{liang2018ultrahigh} introduced a metalens featuring both a high numerical aperture (NA) and high transmission within the visible spectrum using crystalline silicon (c-Si). However, the materials used for these metalenses, such as c-Si and TiO$_2$ for creating high-quality metalenses require complex, time-consuming, and costly manufacturing techniques. These complicated methods make it difficult to produce these metalenses in large quantities, limiting their widespread use. Silicon-rich Nitride (SRN) emerged as a favored material in the development of metasurfaces due to its unique optical properties, compatibility with existing semiconductor manufacturing processes (CMOS), straightforward fabrication process, and stability. Notably, the recent studies \cite{ye2020metalens, goldberg2023silicon}, highlighted the advantages of SRN in creating artificial focus patterns and Huygens meta hologram. This was further supported by the research of our group \cite{ye2019silicon}, in which we demonstrated the SRN-based subwavelength grating metalens and the application of SRN-based metalens in the visible wavelength. In the present study, we concentrate on the crucial role of pitch size in augmenting the efficiency of metalenses and how SRN facilitates the attainment of small pitch sizes, compared to conventional TiO$_2$-based systems\cite{ye2019linear, qian2022broadband}. This is achieved by leveraging the fine-tuning of the composite material's response to P-polarized light using effective medium theory (EMT) and superior refractive index properties of SRN, which allow for a more compact nanocell structure without compromising the phase delay control across the full \(2\pi\) range. utilizing this superiority of SRN and reducing the pitch size, we have proposed two distinct designs. The first metalens (Mk1), with a NA of \(0.9\), achieved a focusing efficiency of \(\%75\) and a Full Width at Half Maximum (FWHM) of \(0.53 \lambda\). The second (Mk2), designed with a higher NA of \(0.99\), achieves an efficiency of \(\%42\) with a FWHM of \(0.48 \lambda\).

\section{Design and Simulation} 

\subsection{Phase Matching and Lens Design}

The development of metalenses is based on arranging tiny nanostructures, smaller than the wavelength of light, in a specific pattern to create a changing phase profile across the lens surface. This phase change is described by the equation:

\begin{equation}
 \phi(x, y) - \phi_0 = \frac{2\pi}{\lambda} \left( \sqrt{x^2 + y^2 + f^2} - f \right) 
\end{equation}

Here, \( \phi(x, y) \) represents the phase shift at a specific point \((x, y)\) on the lens, with \( \phi_0 \) being the initial phase shift at the center. The symbol \( \lambda \) stands for the wavelength of incoming light, and \( f \) denotes the lens's focal length. This equation is fundamental in designing metalenses, allowing for precise control over light by altering its phase across the lens. By incorporating materials with a high refractive index into carefully spaced segments across the lens, it's possible to adjust the effective index at each point, thus controlling the optical response in terms of both light transmission and phase shift.

To cover a full range of phase shifts (a full \( 2\pi \) cycle), materials like TiO2 and GaN are used because of their high refractive properties. The production process, which includes techniques such as atomic layer deposition (ALD) or metal-organic chemical vapor deposition (MOCVD), can be time-consuming and costly but is crucial for achieving the desired lens performance. Reducing the pitch size to minimize the effects of space discretization on the light wavefront can improve the metalens further. This effort aims to reduce losses from scattering or undesired resonances. The effort to make pitch sizes smaller to improve focus is held back by current technology in nano-fabrication, which sets the smallest size for these parts that change the phase. Thus, there's a limit to how small these pitches can be made for a given material and wavelength, which is necessary to achieve a full \( 2\pi \) phase shift range.

\subsection{Enhancing Metalens Focus Efficiency through Pitch Size Reduction and High Numerical Aperture}

\begin{figure}[h!]
    \centering
 
     {\includegraphics[width=\linewidth]{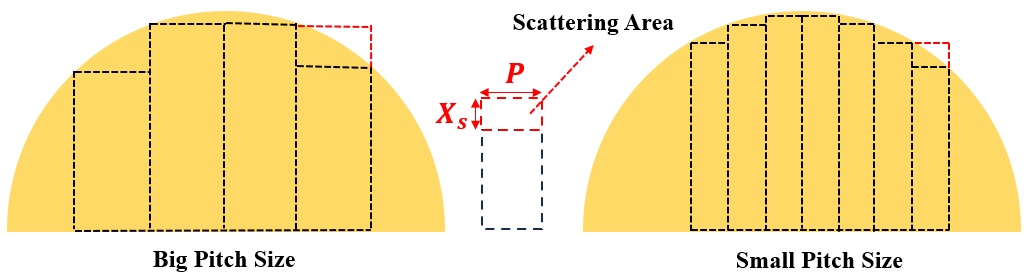}}

    \caption{\small Comparison of metalens nanostructures illustrating the impact of pitch size on the scattering area, where \( P \) represents the pitch and \( Xs \) denotes the optical path difference in adjacent periods.}
    \label{fig:scattering}
\end{figure}

In metalens design, the efficiency of focusing light is fundamentally linked to the precision of phase manipulation across the lens surface. A key factor in this process is the pitch size, represented by \( P \), which is the period or center-to-center distance between adjacent nanostructures. This pitch size directly influences the scattering area, given by the product \( P \times Xs \), where \( Xs \) is the optical path difference between adjacent periods, as depicted in Fig. \ref{fig:scattering}.

The optical path difference \( Xs \) can be mathematically represented as:

\begin{equation}
 Xs = \sqrt{(x + P)^2 + f^2} - \sqrt{x^2 + f^2}/n 
\end{equation}
Here, \( x \) represents the position on the metalens, \( f \) denotes the focal length, and \( n \) is the refractive index of the medium. Decreasing the pitch size \( P \) reduces the scattering area, which allows for more precise phase control and consequently enhances the focusing efficiency of the metalens. The relationship between the pitch size \( P \) and the focusing efficiency is further informed by the Rayleigh-Gans approximation. This approximation is appropriate when the scatterer size is significantly smaller than the wavelength of light divided by the absolute difference in refractive indices \( |n - 1| \):

\begin{equation}
 d \ll \frac{\lambda}{|n - 1|} 
 \end{equation}

The dimension \( d \), associated with \( P \), indicates that as \( P \) is minimized, each nanostructure acts increasingly like an ideal point scatterer, reducing diffraction and aberration effects and thus improving focus efficiency following the Rayleigh-Gans approximation.

\begin{figure}[h!]
    \centering
    \begin{minipage}[b]{0.49\linewidth}
        \centering
        \subcaptionbox{}{\includegraphics[width=\linewidth]{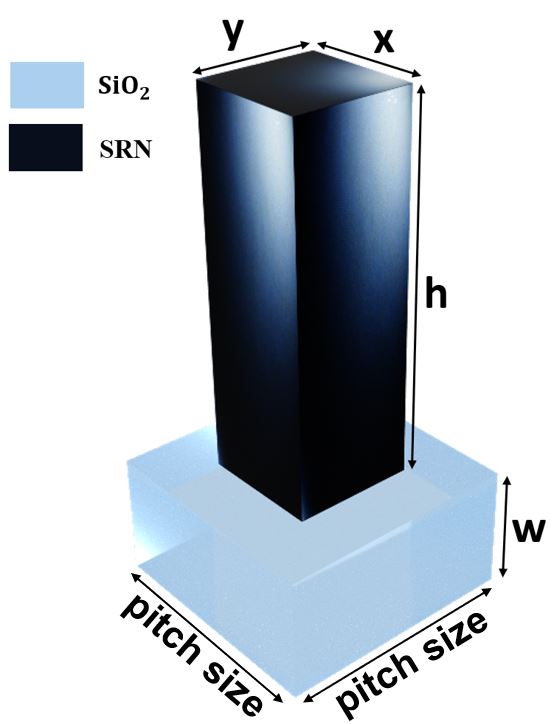}}
    \end{minipage}
    \begin{minipage}[b]{0.45\linewidth}
        \centering
        \subcaptionbox{}{\includegraphics[width=\linewidth]{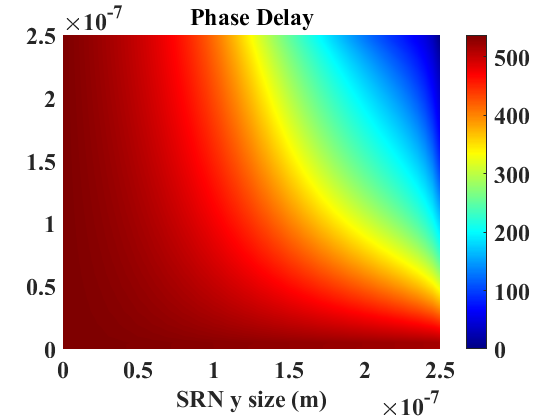}}
        \subcaptionbox{}{\includegraphics[width=\linewidth]{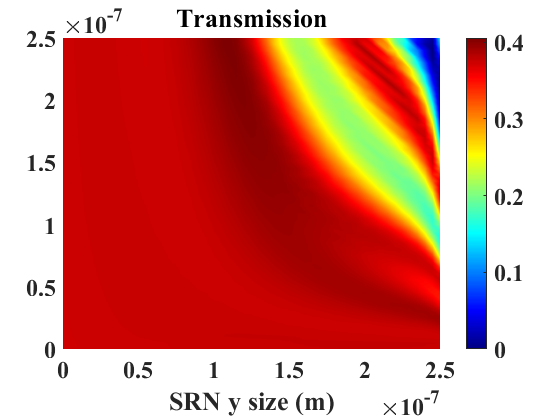}}
    \end{minipage}
    \caption{\small (a) Nano cell dimensions and materials, (b) nanocell's phase shift, and (c) transmission as a function of x and y values.}
    \label{fig:phaseandtrasnmission}
\end{figure}

Additionally, The nano cell's vertical dimension is established through the deposition of an SRN layer, while its lateral dimensions, specifically in the x and y axes, are refined to optimize phase shifts and transmission characteristics (Fig. \ref{fig:phaseandtrasnmission}). To identify the limitations on feature size imposed by fabrication capabilities, we employed equation \(h_{\min}\) to determine the minimal height that allows for effective phase delay control over a \(2\pi\) range, under extreme conditions with or without dielectric presence. Due to the limitations of fabrication processes, which include feature size limitations, an increase in height is required. Thus, a height of 600 nm, corresponding to 1.52 times the minimum height, was selected for the SRN layer to ensure adequate phase management.

\begin{equation}
h_{\min} = \frac{\lambda_0}{n_{\text{dielectric}} - n_{\text{air}}}
\end{equation}

\begin{figure}[htbp]
\centering

% Pair 1: Mk3 and Mk8 layouts

\includegraphics[width=1\linewidth]{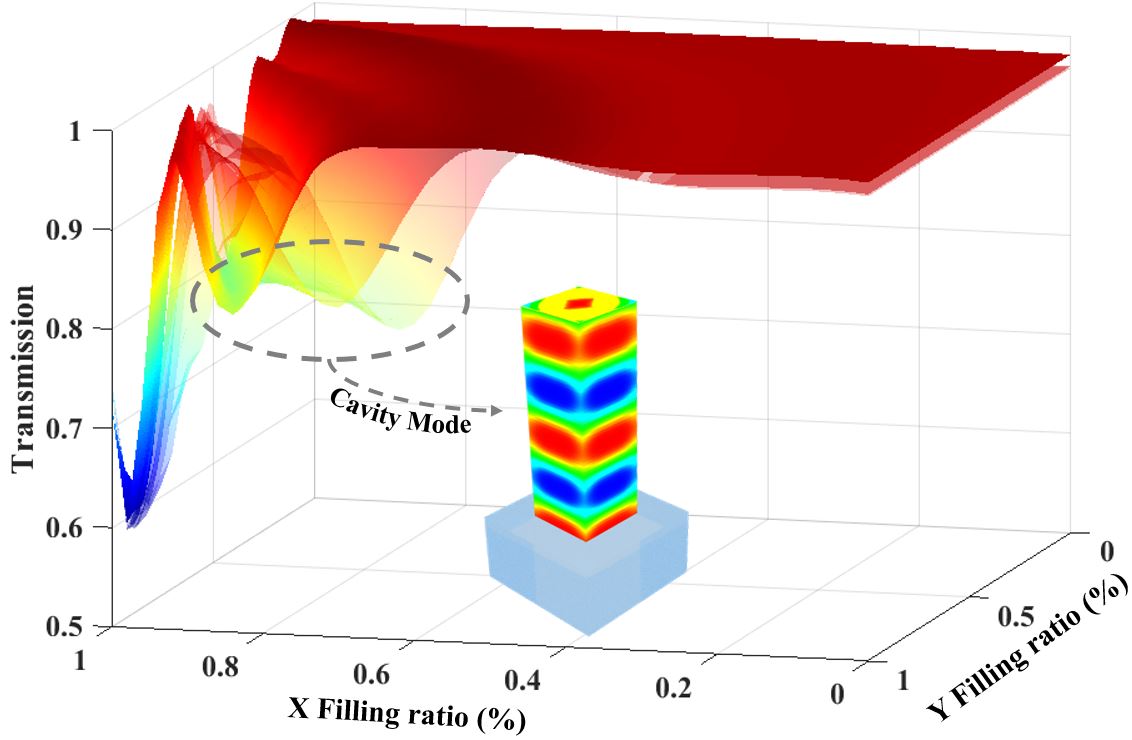}

\label{fig:mk3}

\hfill

\caption{\small Cavity mode within a nanocell, and transmission variation as a function of x and y values of the nanocell.}
\label{fig:combined resonance}
\end{figure}

Furthermore, The equation \(P_{\min}\) was utilized to determine the minimal theoretical feature size necessary to prevent resonant modes within the medium, which would diminish transmission efficiency (Fig. \ref{fig:combined resonance}). \ref{fig:combined resonance} (a) Shows the electric field distribution within a nano cell. The resonant mode indicates high-intensity regions where light is confined, corresponding to the pitch size limitation that can affect overall light transmission. Fig. \ref{fig:combined resonance} (b) illustrates the impact of unit cell parameters on light transmission through a nano cell. The plot shows a decrease in transmission with certain x and y values, highlighting the importance of pitch size in avoiding resonance modes that reduce transmission. Although a reduction in pitch size could theoretically diminish such modes, simulation results indicate that to achieve a \(2\pi\) phase delay, the fill ratio must be maintained between 0.2 and 0.8. Taking fabrication capability into account, we established a pitch size limit of 250 nm. In extreme scenarios, for instance, the dielectric features within a unit cell would range from 50 nm to 200 nm for full and zero phase delays, respectively.

\begin{equation}
P_{\min} = \frac{\lambda_0}{2 n_{\text{dielectric}}}
\end{equation}

\begin{figure}[htbp]
\centering
\begin{subfigure}{1\linewidth}
\includegraphics[width=\linewidth]{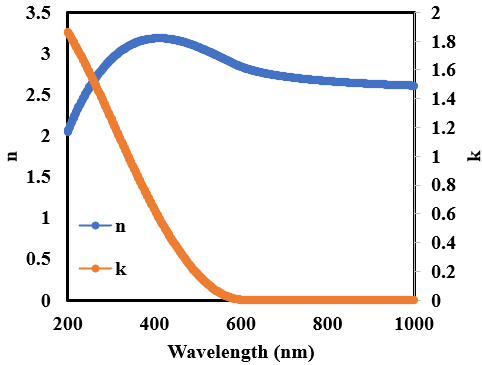}
\caption{} % Add subcaption for (b)
\label{fig:mk8}
\end{subfigure}

% Pair 1: Mk3 and Mk8 layouts
\begin{subfigure}{0.495\linewidth}
\includegraphics[width=\linewidth]{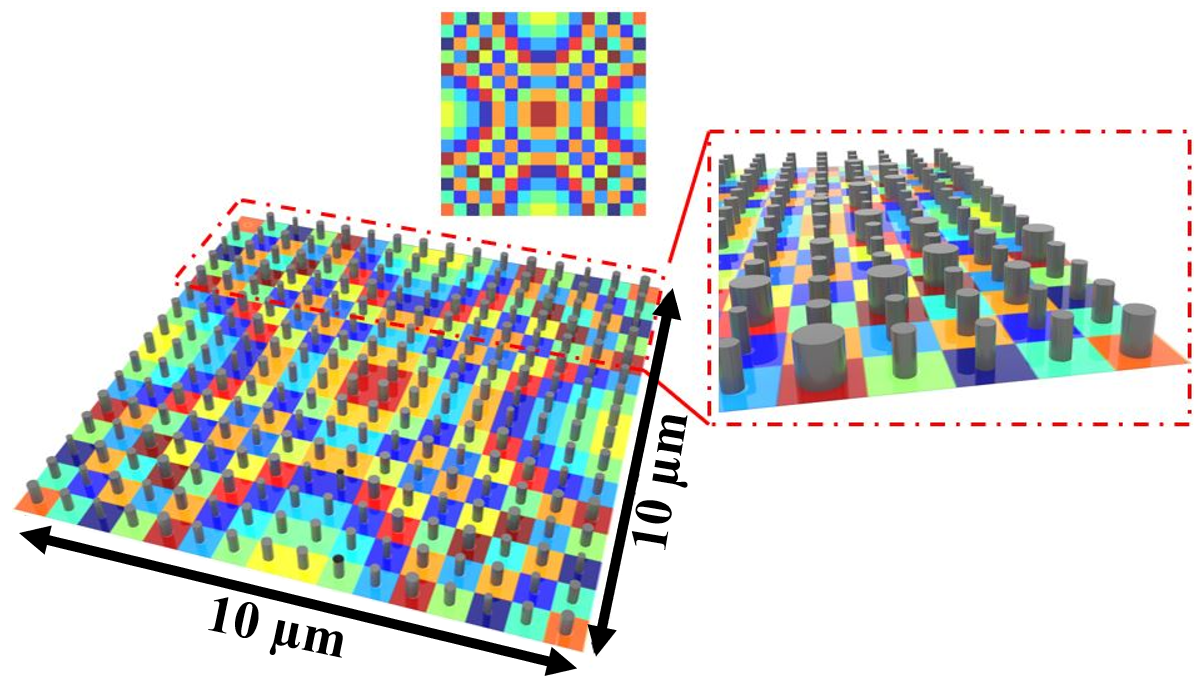}
\caption{} % Add subcaption for (a)
\label{fig:mk3}
\end{subfigure}
\hfill
\begin{subfigure}{0.495\linewidth}
\includegraphics[width=\linewidth]{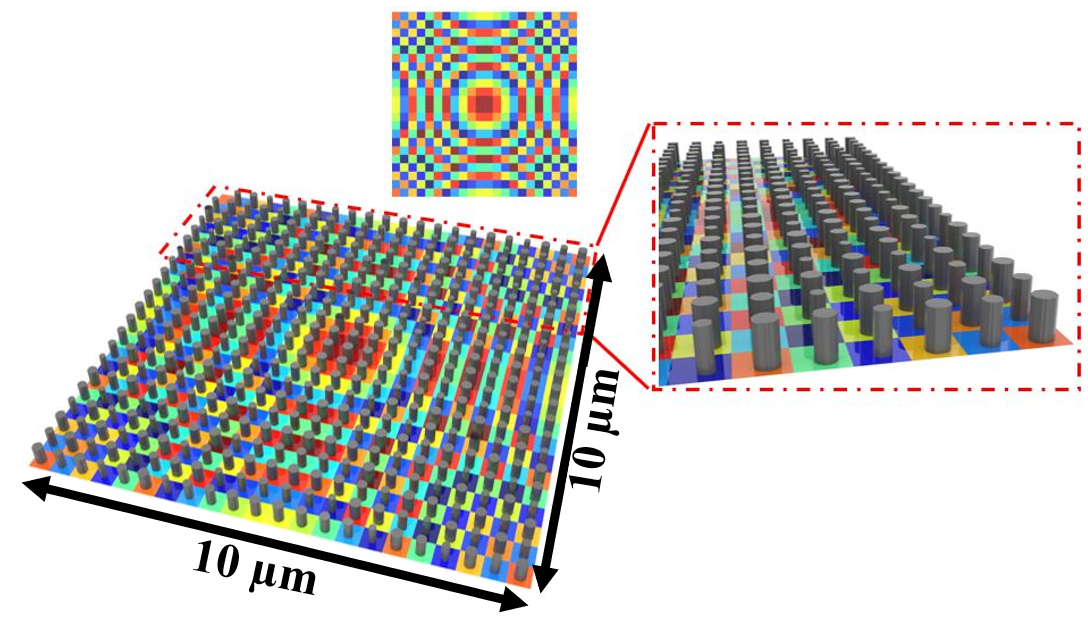}
\caption{} % Add subcaption for (b)
\label{fig:mk8}
\end{subfigure}

\caption{\small (a) SRN n, k values at different wavelengths measured using Woollam Ellipsometer. The phase profile through the employment of two distinct pitch sizes (b) bigger pitch size, and (c) smaller pitch size. (For the purpose of demonstration, the selected pitch size is enlarged.)}
\label{fig:diff-pitch}
\end{figure}

In our research, we utilized Plasma Enhanced Chemical Vapor Deposition (PECVD) to synthesize SRN layers, carefully adjusting the ratio of SiH$_4$ and NH$_3$ gases. By fine-tuning the gas composition to $80\%$ SiH$_4$ and $11\%$ NH$_3$ and operating at a temperature of $350$ degrees Celsius under $3$ torr pressure, we achieved an SRN layer with a thickness of $600$ nm. This precise control allowed us to obtain a refractive index of $2.74$ at a wavelength of $685$ nm, as confirmed by Woollam ellipsometer measurements presented in Fig. \ref{fig:diff-pitch} (a). This high refractive index, is comparable to that of materials like TiO$_2$ or GaN used in similar studies, enabled us to achieve a remarkably small pitch size of $250$ nm. This reduction in pitch size \ref{fig:diff-pitch} (a) and (b), critical for our metalens design, illustrates the potential of our method and developed material in minimizing spatial discretization effects and enhancing the overall efficiency of metalenses, setting a new benchmark for bandwidth applications beyond $600$ nm.

\subsection{Proposed Metalenses Performance Evaluation}

\begin{figure}[htbp]
\centering

% Pair 1: Mk3 and Mk8 layouts
\begin{subfigure}{0.49\linewidth}
\includegraphics[width=\linewidth]{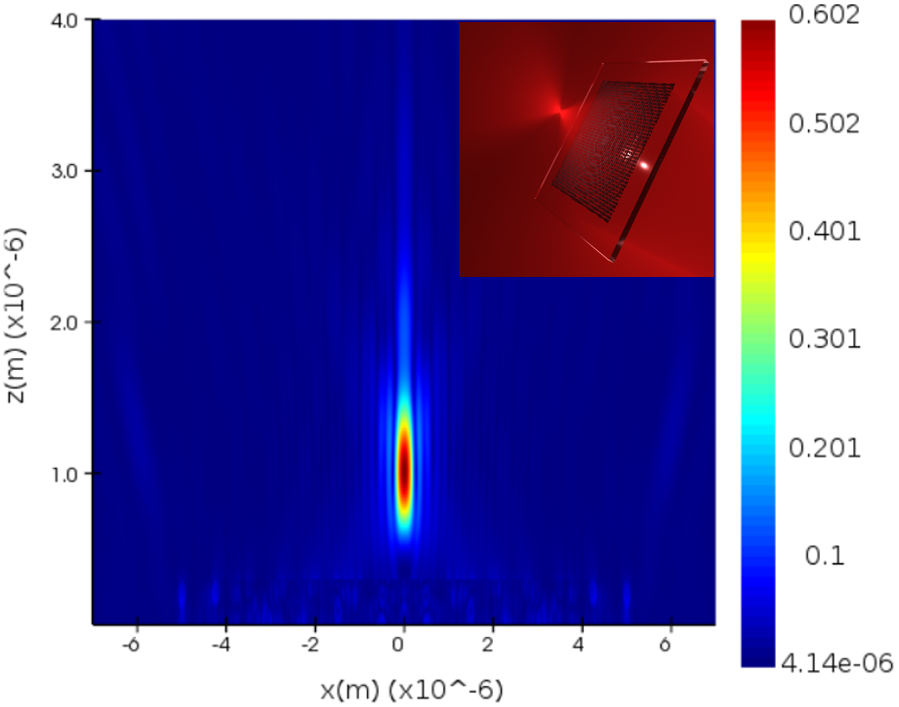}
\caption{} % Add subcaption for (a)
\label{fig:mk3}
\end{subfigure}
\hfill
\begin{subfigure}{0.49\linewidth}
\includegraphics[width=\linewidth]{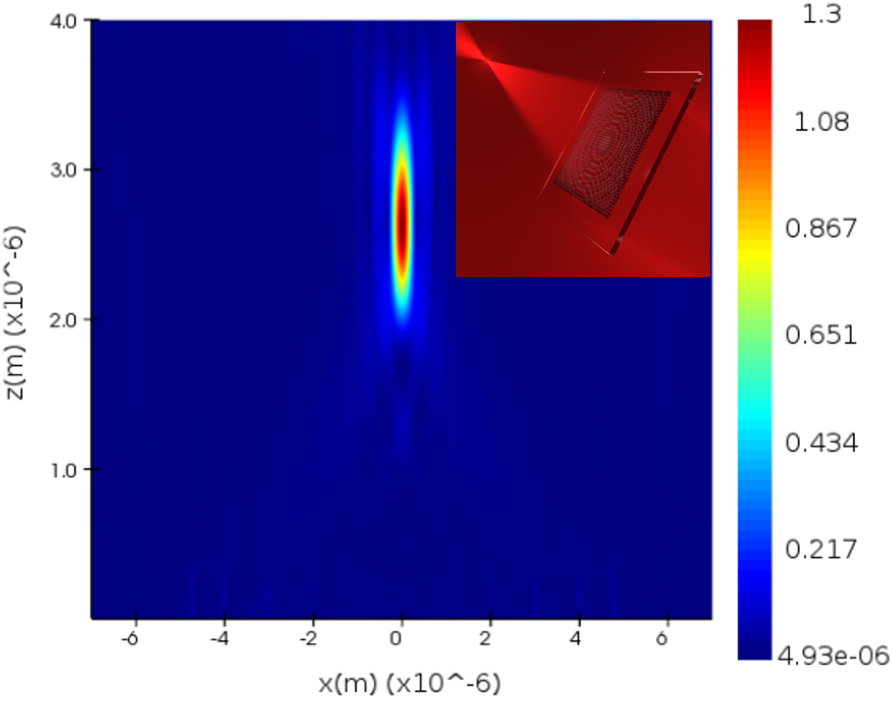}
\caption{} % Add subcaption for (b)
\label{fig:mk8}
\end{subfigure}

\caption{\small E-field distribution of the metalenses under an incident wavelength of 685 nm of the (a) Mk1 and (b) Mk2 micro grating lens structures, along with their 3D visual illustrations .}
\label{fig:3d and e-field}
\end{figure}

Our simulation process utilized the Finite Difference Time Domain (FDTD) method. A P-polarized light source was employed to specifically target the characteristics of our metalens design. The PML boundaries were selected as the boundaries of the simulation. Incident light at a wavelength of 685 nm was selected as the light source. The corresponding n, k values of SRN at 685 were obtained from our experimental investigation of SRN using Woollam ellipsometer Fig. \ref{fig:diff-pitch} (a). The performance evaluation of the metalenses focused on key metrics such as focusing efficiency, NA, and FWHM. Focusing efficiency is calculated by integrating Poynting vectors over the incident area and at the focal spot, defined by three times the FWHM of the intensity. The efficiency is the ratio of these two integrations, reflecting the lens's ability to concentrate light. The Mk1 lens, with a high NA of 0.9, demonstrated a remarkable 75\% focusing efficiency, whereas, the Mk2 lens, featuring a higher NA of 0.99, achieved a 42\% efficiency, highlighting its precision in light focusing despite the higher NA. Resolution was assessed by the FWHM, with the Mk1 and Mk2 lenses achieving $0.53\lambda$ and $0.48\lambda$, respectively. These metrics underscore the advanced performance of our high-NA metalenses in efficiently focusing light and providing sharp resolution. Fig. \ref{fig:3d and e-field} offers a 3D depiction of both Mk1 (part a) and Mk2 (part b) metalenses, alongside the side view of the E-field distribution at the focal point within these metalenses under an incident wavelength of 685 nm.

%\section{Proposed Fabrication Method}
%This section outlines a method to fabricate polarization-sensitive metalenses. Starting with a 4-inch, 500 µm thick glass wafer as the base, a 600 nm Silicon Rich Nitride (SRN) layer is deposited to form the lens material. Over this, a 300 nm Silicon Dioxide (SiO$_2$) layer is added to serve as an etching mask. Both layers are applied using Plasma Enhanced Chemical Vapor Deposition (PECVD) for uniformity.

%A 200 nm ZEP520A photoresist layer is spin-coated on the SiO$_2$ and patterned via E-beam lithography, defining the lens design. The pattern is etched into the SiO$_2$ using reactive ion etching (RIE), and then into the SRN, forming the metalens. The SiO$_2$ mask is finally removed with hydrofluoric acid. A two-step RIE process is necessary due to the photoresist's limited etching selectivity against SRN and the design's fine 48 nm features.

\section{Comparision}

Table \ref{table:metalenses} compares various metalenses, including our proposed designs Mk1 and Mk2, to recently proposed metalenses. Mk1, known for its high efficiency with an NA of $0.9$ and $75\%$ efficiency, offers a balance between performance and energy conservation, featuring an FWHM of $0.53\lambda$ and a $2.33\ \mu\text{m}$ focal length at $685\ \text{nm}$. This makes it ideal for quality imaging in applications requiring an extended focus, with the added benefit of CMOS compatibility and a straightforward fabrication process due to SRN construction. Mk2, also based on SRN, provides a higher NA of $0.99$, showcasing a better light-gathering and resolving power for high-resolution imaging over Mk1, despite a lower efficiency of $43\%$. It also offers an FWHM of $0.48\lambda$ and compact focal length of $0.709\ \mu\text{m}$ at $685\ \text{nm}$.

The design by Lu et al., employing aluminum, presents an NA of 0.44 and an efficiency of 69\% \cite{lu2018broadband}. Its FWHM of 540 nm and a focal length of 3.69 $\mu$m at 550 nm suggest a broader range of application but potentially less precision compared to the compact and focused Mk1 and Mk2. Kim's proposed metalens, combining Silicon and Germanium, features an NA of 0.5 and a lower efficiency of 27\% with its FWHM of 750 nm and extended focal length of 20.78 $\mu$m at 700 nm \cite{kim2022design}. Another design by Zhu et al., made of Silicon, offers an NA of 0.47 and a high efficiency of approximately 77.7\% \cite{zhu2023ultra}. It provides a broader focus with its FWHM of approximately 800 nm and a focal length of 9.5 $\mu$m at 720 nm, differing from the precise focus of Mk1 and Mk2. Chung et al. proposed a design, utilizing Titanium Dioxide, that matches Mk1 in NA but with focusing efficiency at approximately 23\% with a focal length of 3 $\mu$m at 700 nm\cite{chung2020high}. Lastly, the Liang et al. designed metalens, fabricated from crystalline Silicon, features a high NA of 0.975 and an efficiency of 58.4\% \cite{liang2018ultrahigh} with the focal length of 105 $\mu$m at 532 nm and an FWHM of 0.207 nm.

Overall, Mk1 and Mk2 demonstrate superior attributes in comparison with other metalenses designs. Mk2 excels in high-resolution imaging with its high NA and tight FWHM, while Mk1's enhanced efficiency and commendable optical properties make it versatile. Notably, both Mk1 and Mk2, designed based on SRN offer a promising, straightforward. and CMOS-compatible fabrication process, facilitating their integration into advanced semiconductor-based devices and photonic integrated circuits. The advancements of these SRN-based metalenses offer high efficiency along with remarkable NA, positioning them at the forefront of metalens design.

\begin{table}[htbp]
\centering
\caption{\bf Comparison of Various Metalenses}
\resizebox{\columnwidth}{!}{
\begin{tabular}{lccccc}
\hline
 & Material & NA & Efficiency (\%) & FWHM & Focal Length ($\mu$m) \\
\hline
Mk1 & SRN & 0.9 & 75 & 0.53$\lambda$ & 2.33 at 0.685 \\
Mk2 & SRN & 0.99 & 43 & 0.48$\lambda$ & 0.709 at 0.685 \\
\cite{lu2018broadband} & Al & 0.44 & 69 & 540 nm & 3.69 at 0.55 \\
\cite{kim2022design} & Si and Ge & 0.5 & 27 & 750 nm & 20.78 at 0.7 \\
\cite{zhu2023ultra} & Si & 0.47 & $\approx$77.7 & $\approx$800 nm & 9.5 at 0.72 \\
\cite{chung2020high} & TiO$_2$ & 0.9 & $\approx$23 & Not spec. & 3 at 0.7 \\
\cite{liang2018ultrahigh} & c-Si & 0.975 & 58.4 & 0.207 nm & 105 at 0.532 \\
\hline
\end{tabular}%
}
\label{table:metalenses}
\end{table}

\section{Conclusion}

In this study, we've highlighted the critical role of pitch size in leveraging the tunable high refractive index of SRN for high NA metalens design. Our investigation into SRN's application has led to the development of Mk1 and Mk2 metalenses, demonstrating how pitch size optimization can significantly enhance metalens focusing efficiency. Mk1, with an NA of 0.9, achieved a 75\% focusing efficiency and a FWHM of $0.53\lambda$, while Mk2, with an NA of 0.99, reached a 42\% efficiency with a FWHM of $0.48\lambda$. These results underscore the profound impact that precise pitch size adjustments can have on improving both NA and FWHM, directly translating to higher lens focusing efficiency and better performance. This focus on pitch size not only advances our understanding of SRN's potential in nanophotonics but also sets a new benchmark for the design of efficient, high-performance metalenses.

\begin{backmatter}

\bmsection{Disclosures} The authors declare no conflicts of interest.

\bmsection{Data availability} Data underlying the results presented in this paper are not publicly available at this time but may be obtained from the authors upon reasonable request.

\end{backmatter}

% Bibliography
\bibliography{ref.bib}

% Full bibliography added automatically for Optics Letters submissions; the following line will simply be ignored if submitting to other journals.
% Note that this extra page will not count against page length

\end{document}